# Network-Based Analysis of Public Transportation Systems in North American Cities

Abbas Masoumzadeh[1]
Department of Electrical Engineering
and Computer Science, York University
Canada
abbasmz@eecs.yorku.ca

Tilemachos Pechlivanoglou[1]
Department of Electrical Engineering
and Computer Science, York University
Canada
tipech@eecs.yorku.ca

## ABSTRACT

A comprehensive data analysis system is implemented for the extraction of information and comparison of North American public transport systems. The system is based on network representations of the transport systems and makes use of a span of metrics and algorithms from the established properties in graph theory to complicated domain specific measurements. Due to nature of big data systems and the requirement of scalability, many heuristic optimizations and approximations have been considered in the system. Integration with other sources of data specially population density maps is also executed in the system. Formal evaluations are done on subcomponents of the system to make sure the approximations have reasonable precision. Results on comparison of four cities, San Francisco, Boston, Toronto and Los Angeles, approves that the big data approach to comparison of public transit systems can successfully reveal the underlying similarities and differences.

• **Human-centered computing** → Visualization

## 1 INTRODUCTION

American Public Transportation Association (APTA) has named Toronto Transit Commission (TTC) as the best public transport agency in North America for 2017. While this award should make Toronto citizens proud, many of them are unsatisfied with the quality and consistency of the services provided by TTC. Objectively analyzing, visualizing and comparing the public transportation systems of some of the largest North American cities is possible in order to clarify this difference of viewpoints.

Thanks to the availability of interconnected and open systems, a variety of data sources on the structure and status of public transportation in North America is available and can be used as the data domain of the current study. By formatting the collected data and adjusting it to the concepts of graph theory, issues and questions regarding public transportation systems are mapped to problems in large graph data analysis, allowing them to be addressed using established methodologies and metrics from the field.

Trying to answer the question "which city in North America has the best public transport system?", could lead to an unending discussion between people with different viewpoints and expectations backed by subjective analysis of one or a few limited sources of data. To address this issue, it is important to divide the question in to a set of sub-questions that can form an objective evaluation of public transportation systems [1, 20]. Consider the following as a few candidate questions:

- Which covers the most area?
- Which offers the most access?
- Which allows the fastest trips?
- Which is most efficient in terms of structure?

It is important to apply well-defined analytics on the vast amount of data and to visualize those metrics in a clear and understandable manner.

The main application of such analysis would be a more consistent urban planning strategy by specifying the issues in each city and pointing to established solutions in cities with clearly better status in those matters. Furthermore, organizations can combine their own sources of information with the results and apply dynamic analysis to better understand the bottlenecks of their systems, such as emergency systems' logistics, delivery services etc.

---

[1] Author names are in alphabetical order.

## 2 PROBLEM DESCRIPTION

### 2.1 Motivation

The goal for defining the transit networks as graph models is to transform the problem of comparing transit systems of two or multiple cities and various concepts that can be defined about them, to the problem of comparing network-based models and the properties of network models. The first step toward the goal is to define how to map the transit system of a single city to a graph.

### 2.2 Network model

For simplicity, the most basic form of the model is presented here. The bus/metro stop stations are defined as the vertices of a graph and whether or not two stops are connected via a bus route or metro rail, defines the existence of an edge between them. More sophisticated network models can be defined by considering directions for the edges, one or multiple types of weights for the edges based on different notions of distance between two stops, including direct distance on map, road distance, average time on public transit distance and etc.

### 2.3 Problem Definition

The inputs of the main problem are data-sources giving access to the data on transit network systems of multiple North American cities and additional data on aligned topics of interest like population density models and the outputs are a set of analysis metrics to be computed for the transit systems of those cities. The computed analysis metrics are presented in a uniform way, to be used for decision making by the user. Analyzing the main problem in a comprehensive manner entails complexity and non-homogeneity. Fortunately, it can be divided into independent sub-problems, each of which can be studied separately. These sub-problems are mainly of three types. First, problems that faced when trying to build a framework for comparing network models. Second, problems of mapping properties of the transit systems to graph and pathfinding problems in an efficient manner to have a platform for analysis of metrics. Third, problems regarding analyzing the metrics in a big data problem with scalable and reasonably precise approaches. Here we introduce a number of more interesting problems, define them formally and discuss our approaches to solve them. Some of the more algorithmic components are provided in more details in section 4.5.

*2.3.1 Problem 1: Graph connectedness.* One of the main obstacles in calculating the intended properties of the network is the graph connectedness. To be more specific, many algorithms that define the properties of a graph model, rely on the assumption of having a connected graph model as input. However, the graph model of a public transit network could easily be disconnected. Consider the case of having two bus stops or a bus stop and a metro station, being near, but people have to walk in between to switch from one to the other. Finding the reason for such disconnections between two nodes of a graph is of course not hard if one plans to study each in the documents and correct it manually in the model. However, since this is a big data project and the approach should be scalable to the public transit network of any city or even other similar concepts in cities, it should be done automatically. This leads to definition of a problem of how to form a connected graph out of two or multiple connected components.

*2.3.1.1 Problem definition.* The simplest form of the problem considers a disconnected graph G(V, E) with homogenous definitions of vertices V and edges E. The input of the problem is a graph, modelling a public transit network, which is most probably not connected. The required output is providing a set of edges E' adding which will have minimum impact on the properties of the network while converting the graph to a connected graph.

*2.3.1.2 Hardness of Problem.* According to Gosh and Boyd [6], a general case of the problem can be modeled with a Boolean problem, solving which requires calculating the second smallest eigenvalue for $\binom{E_c}{k}$ Laplacian matrices, where $E_c$ is the set of candidate edges to choose from and k is the intended number of edges.

*2.3.1.3 Approach.* In order to avoid costly calculations in the project and keep the solutions scalable to even larger datasets, solving the actual problem is bypassed by doing a conversion in the representation of the model. First the distance between each two non-neighbor stops is calculated and if they are nearer than a threshold distance, then they are merged without losing the information regarding their routes and since their distance is less than a small threshold, the new node is placed on the middle point of the direct path between the initial nodes. This way not only the problem of connectivity will be solved but also the model will be ready for calculation of parameters, such as finding the shortest path between two stops, which would have been wrongly calculated in the previous state.

*2.3.2 Problem 2: Public transit area based coverage.* One of the metrics that naturally translates to development rate of a transit system in a city is how well the area of city is covered by the transit system. Considering that a 400 meter (5 minutes) walk from home or office to a transit stop should be pleasant [10], public transit area based coverage can be expressed as the set of two values, the number of stops that on average a person can find starting from habitable areas of the city, and the average distance to the nearest stop.

*2.3.2.1 Problem definition.* The input of the problem is the data on the transit system of a city, and the output of the problem is the set of two area based coverage values explained above.

*2.3.2.2 Hardness of the problem.* The exact answer to the problem requires an exhaustive approach over all possible coordinates in the map of the city, which if discritisized with one square meter precision, for a city like Toronto with more than 630 $Km^2$ area would be unreasonably time consuming.

*2.3.2.3 Approach.* In order to solve this problem with a low computational complexity and reasonable precision, two approximations were considered. First, the starting positions were limited to 10,000 randomly sampled positions over a uniform distribution. Second, habitable areas of a city are not farther than 800m from at least one station or stop in the transit network. The detailed procedure steps are discussed in section 4.5.2.

*2.3.3 Problem 3: Population density map.* Some analysis metrics requires the knowledge on the population density in specific locations. Statistical data on US zip codes and Toronto neighborhoods is publicly available.

*2.3.3.1 Problem definition.* Input of the problem is the data sources mentioned above for a specific city. Output of the problem is a population density map of neighborhoods and matching of transit system stops based on their coordinates.

*2.3.3.2 Hardness of the problem.* The complexity of the problem arises from the shapes of the neighborhoods not being simple geometric shapes and requiring exhaustive search over the map of the neighborhoods.

*2.3.3.3 Approach.* An approximation on the shape of the neighborhoods is made as them being squares with their centers located at the coordinate of the zip code/ward. This way, calculating the population density in each neighborhood would be trivial and fast.

*2.3.4 Problem 4: Public transit population based coverage.* This problem is similar to problem 2, with the difference on starting points being defined according to the population density.

*2.3.4.1 Problem definition.* The input of the problem is the data on the transit system of a city, and the output of the problem is the set of two population based coverage parameters.

*2.3.4.2 Hardness of the problem.* The problem's hardness is also similar to problem 2 with the addition of calculation of population density.

*2.3.4.3 Approach.* A similar approach to the one provided by problem 2 is used. The only difference is that randomly sampled points are over a population density rather than a uniform distribution. The detailed procedure steps are discussed in section 4.5.3.

*2.3.5 Problem 5: Minimum edge coloring for a path.* This problem arises when trying to calculate the least amount of time to go from a specific point to another specific point in the map of the city, using the public transit system.

*2.3.5.1 Problem definition.* This problem can be mapped to a graph $G(V, E)$ as finding the shortest path from a node $A \in V$ to another node $B \in V$. Edges have individual weights and they also belong to color classes, each of which has a weight which is only in effect once in the path if an edge belonging to the color class belongs to the path. Shortest path is defined as the path minimizing the sum of total weight of its edges and weight of edge classes existing in the path.

*2.3.5.2 Hardness of the problem.* If there were at most k color classes in a path containing $|V|$ nodes and $|E|$ edges, knowing that Dijkstra algorithm can be solved in $O((|V|+|E|)\log(|V|))$ [5], combining it with the color problem, since in the worst case each edge could take k different colors, the problem can be solved in $O((|V|+k|E|)\log(|V|))$.

*2.3.5.3 Approach.* In order to solve the problem faster we first ran the Dijkstra algorithm not caring about the colors (routes) and getting the shortest weight (travel-time) path. Then we applied the minimum edge coloring algorithm, and we applied a naïve static weight to counter the effects of the greedy pathfinding algorithm. The details of the algorithm are discussed in section 4.5.4.

*2.3.6 Problem 6: Public transit access to points of interest.* This problem is considered as a generalized metric through which users can assess the access to their points of interest, for example the shopping centers in a city. Two parameters, the minimum distance and minimum time required to reach to the nearest point of interest on average, starting from any point in the city, using the public transit system.

*2.3.6.1 Problem definition.* The input of the problem besides the transit system data of the city is also a set of points of interest. The output of the problem is a set of two parameters describing the access to the set of points of interest. Which are explained above.

*2.3.6.2 Hardness of the problem.* The hardness of the exact problem can be easier to understand by considering the exact problem as a combination of problems 5 multiplied by the number of points of interest and problem 4 multiplied by the number of points in a city to start from.

*2.3.6.3 Approach.* The solution of the problem will also be a similar combination of the problems 4 and 5. More details on the algorithm can be found in section 4.5.5.

## 3 RELATED WORKS

Derrible and Kennedy [2], suggested metro transit networks to be modeled using graphs. They have characterized metro networks' state, form and structure in terms of graph theoretic measures. Farahani et al. [3], reviews many analysis metrics that have been presented in the literature and classifies their algorithmic methods.

Public transit networks have also been addressed in terms of connectivity in the literature before. Yang et al. [19] studies connectivity and tries to standardize it in the concept of public transit networks by suggesting the consideration of different ranges of connectivity as indices for evaluation of the maturity of a public transit network. Mishra et al, [9] discusses performance indicators for public transit connectivity which leads to methodologies for calculating connectivity for nodes, lines, transfer center and a region in the network.

Zielstra and Hochmair [22], comparatively studied penetration access to transit system stations in US and German cities. Saghapour et al. [12] studied public transport accessibility using GIS techniques and population density, to impartially evaluate accessibility levels in metropolitan areas. Widener et al. [18] studies the accessibility to supermarkets using public transit and compares it to previous models of the same concept for car drivers.

Van Oort et al. [16] considered the unreliability of the network to make a more accurate demand model.

Farber and Fu [4] introduced the public transit travel time cube containing the shortest path transit travel times between sets of origins and destinations in the city, at all times of day. It is used to compare how transit travel times and accessibility have changed in response to network and service modifications.

Goal of transfer optimization has been another point of view in studies on modelling public transit networks. Shafahi and khani [14] provided a genetic algorithm to approximately minimize a transfer waiting time function for a large bus network model.

Salonen and Toivonen [13] compared transfer time between private car and bus network models and found similarities and differences in general between models used for each.

Zhang et al. [21] provided a comprehensive evaluation from multiple perspectives. Wei et al. [17] proposed a new method for evaluating the overall performance of public transit services via a combination of data envelopment analysis (DEA), Geographic Information System (GIS), and multi-objective spatial optimization techniques to on measures of operational efficiency and access equity.

## 4 METHODOLOGY

### 4.1 Implementation Iterations and Steps

To ensure that meaningful results can be produced while adhering to the data analysis requirements and milestones, the project follows an iterative process with growing complexity. In each iteration one or more of the steps, containing data collection, processing, analysis and visualization are done.

The key implementation milestones achieved are:
1) Primary visualization of the transportation network
    a) Retrieval of data regarding the targeted cities
    b) Creation of a database for routes, stops and locations
    c) Visualization of the results on geographical city maps
2) Finding properties of networks
    a) Computation of the network properties
    b) Visual depiction of those properties
3) Comparing results for various cities
    a) Provisions for repetition of analysis for various cities
    b) Visualizations to highlight city highs and lows
4) Combination with other data sources and advanced queries
    a) Integration with population density maps
    b) Computation of more complicated queries

Many graph properties have been examined, in terms of applicability to the problem and computational requirements. Patterns were also visualized in an easy to interpret way [8].

### 4.2 Implementation Challenges and Solutions

Main challenges faced during the project were:
- The combination of data from different sources and
- The comparison of data from multiple cities with fundamentally different underlying structure
- The need to answer some computationally complex queries

To resolve these challenges, various methods were used:
- For the first challenge, available data sources with unified matching keys were merged and used.
- The second challenge is a descriptive issue and needs to be resolved on a case by case basis.
- For the third challenge some approximations were made to make sure the solutions will be appropriate considering big data environment

### 4.3 Data Analytics Architecture and Process

The system has two main components, builder and visualizer. The implementation is done using Python and intermediate data results are stored in files to decrease the dependency of the system in case of requirement to migrate to other environments like distributed systems.

The builder is responsible for communicating with the public transport systems' API, retrieving lists of stops, routes, and connections based on those routes. After parsing the data, it does pre-processing such as removing isolated/obsolete stops and finalizes the static network structure.

The visualizer is responsible for calculating the analysis metrics, drawing and visualizing the model.

### 4.4 Analysis Metrics

The required data analysis metrics for this project are based on both statistics and graph theory. They range from simpler analytics like the computation of average delay for a route to medium difficulty queries like computing the shortest route between two stops, to even computationally hard algorithms in graph theory like comparing networks of different cities simultaneously. It is worth mentioning that the analysis techniques and metrics are considered adequate to address the targeted questions, because they can be directly mapped to real world characteristics of a public transportation system. Few examples are: average trip duration, longest possible distance and time, etc.

### 4.5 Specific Algorithmic Components

Duo to the comprehensiveness of the project, multiple algorithms and approaches devised from literature will be used. Since it is not suggested to re-implement the algorithms that are already available in form of libraries implemented by the experts of a field, such algorithms are used whenever possible. To mention a few, consider average shortest path and average clustering coefficient in a graph. For such metrics only additional normalizations may have been considered where seemed useful in the domain of public transit networks.

Here we present a few algorithms which implementation from scratch was required. References are provided when used to address scientific concepts of the algorithm.

*4.5.1 Cosine-Haversine formula.* The data collected on public transit from Nextbus, contained location data in latitude-longitude format. In order to calculate distances in meters between each pair of locations, we used the Cosine-Harvesine formula [15].

$$d = 2r \arcsin\left(\sqrt{hav(\varphi_2 - \varphi_1) + \cos(\varphi_1)\cos(\varphi_2) hav\left(\frac{\lambda_2 - \lambda_1}{2}\right)}\right)$$

Where d is the distance between the two points, r is the radius of Earth at that latitude and the location's height above the sea, $\varphi_i$ and $\lambda_i$ are respectively the latitudes and longitudes of the two points and hav is the Harvesine function:

$$hav(\theta) = sin^2\left(\frac{\theta}{2}\right)$$

The formula is approximated based on a spherical Earth model. However, it is accepted to be accurate for short distances of a few kilometers [11]. The radius of Earth is approximated using the following formula:

$$r = \sqrt{a/b}$$

$$a = [(equ^2 \times cos(lat))^2 + (pol^2 \times sin(lat))^2]$$

$$b = [(equ^2 \times cos(lat)) + (pol \times sin(lat))^2]$$

Where equ and pol are respectively constant approximations of Earth's radius at equator (6378.137 km) and poles (6356.752).

*4.5.2 Public transit area based coverage.* In order to compute the coverage of the map of a city by the public transit network, the following procedure is implemented.

(1) First a threshold of 400 meters is fixed.

(2) 10,000 points are randomly sampled from a uniform distribution of latitudes & longitudes, excluding locations outside the transit system's service area (more than 800 meters from a stop).

(3) For each sampled point, stops that are in a square with side 400 meters away centered around the point are identified.

(4) Then the true distances between the sampled point and the found stops are calculated and the stops more than 400 meters away are removed.

(5) Shortest distance and number of stops found are stored.

(6) The output of the procedure is the average of each of the two values over all sampled points.

*4.5.3 Public transit population based coverage.* This algorithm borrows the algorithmic components of the previous method (4.5.2) and uses the population density map, discussed earlier in section 2.3.3. At the first stage, instead of sampling random points from the whole city with a uniform distribution, it randomly chooses a neighborhood, where the probability of choosing a neighborhood depends on its ratio of population to the whole city population. Next, it chooses a random point from that neighborhood using a uniform distribution. It follows by reusing the public transit network area based coverage algorithm (section 4.5.2) to finalize the metric result.

*4.5.4 Minimum edge coloring for a path.* For this calculation, we maintain a list of candidate edge colors (routes) that are valid up until the last change (transfer). Then we follow these steps:

(1) Initially, the routes of the first connection are selected as candidate routes

(2) The path edges are iterated through in sequence

(3) For every connection, the candidate colors (routes) are iterated through

(4) If the candidate is not present in the list of colors (routes) of the edges, remove it from the list of candidates

(5) If there are no more candidates left, it's not possible to traverse the last leg in a single color (route), so increase the count of transfers required

(6) After reaching the end of the path edges, return the final number of changes (transfers)

*4.5.5 Public transit access to points of interest.* An interesting measure to have for each city is how well a set of points of interest are accessed through the public transit network. To clarify, "how well" here is approximated by considering on average from randomly sampled points in the city based on its population density, a set of 1,000 starting points and calculating the average time to reach the nearest point from the set of points of interests. The algorithm is implemented as a generalized function so that depending on the provided dataset, different concepts can be measured with the same function. The algorithm reuses some portions of algorithms 4.5.2 and 4.5.3 based on the required output type to be area based or population based. It uses the mentioned algorithms to calculate both starting points and ending points. It also uses algorithm 4.5.4 on every sampled point to find its minimum distance to nearest point of interest.

## 4.6 List of Tools and Libraries Used

4.6.1. "Networkx" for storing and manipulating graphs [7].

4.6.2. "Google Maps" API and "OSRM" API for road distance calculations and evaluations.

4.6.3. "Matplotlib" to draw and visualize the graph model.

4.6.4. "Element Tree" for XML parsing.

## 5 EVALUATION

## 5.1 Evaluation Requirements

In this project, evaluation is done in high-level and low-level approaches. High-level, considers the whole framework and evaluates whether as a big data framework it can achieve its goals. Low-level, evaluates its subcomponents regarding analysis of metrics and answers to the question of their precision.

Based on the nature of the analysis that is done in this project, for high-level approach, three different requirements are considered that need to be achieved. Also, to evaluate the achievement of these requirements, some experiments are considered.

- First, the system should be able to work with cleaned and formatted data from various cities, using the NextBus-compliant public transport API for that city. As a measure of consistency, beside the chosen cities for visualization, data from city of Chicago through another framework is also fed into the framework and robustness of the framework was tested.
- Second, the system should be able to work with arbitrary scales of urban data. The three cities of San Francisco, Toronto and Los Angeles were chosen with goal of illustrating the capabilities of the framework with smaller to larger cities. If there would be a requirement to scale and accommodate very large inputs, the effort to modify the implementation into a distributed architecture using Apache Spark would be minimal.
- Third, the visualizations produced by the framework should be able to perform smoothly without the need of manual enhancements on a case by case basis. This is an inherent property of the visualizations considered for the system and is tested on multiple data-sources.

In order to formally evaluate the system using the low-level approach, all the mentioned algorithms in section 4.5 of the report should be examined. However, for algorithms 4.5.2, 4.5.3 and 4.5.5, 10,000 points are randomly sampled to calculate the metrics. Since the ground truth have its roots in a discretization problem, the only way to calculate the ground truth is to sample more points, for which the computational time would be higher and in fact if it would be feasible to sample more points, the same could be done for the algorithms too. So, there is no reasonable way to have a ground truth for them. Therefore, there is no way to formally evaluate them. For the other two algorithms, in order to make sure the algorithms are working correctly, the performance of the algorithms is compared to performance of Google maps API as the ground truth. The expectation is not to be exact same answer but to be in a reasonable neighborhood so that there would be a hope of convergence in case of enlarging the quantity of the test according to the law of large numbers. The details of test protocols and results, follows:

- For algorithm 4.5.1, considering results from Google Maps as ground truth, 3 random pair of stops from 3 cities were chosen and the calculated distance of the algorithm were compared. The error rate was calculated on average as 0.0000069%.
- For algorithm 4.5.4, 3 random pairs of points from 3 cities were chosen and the minimum number of transfers were calculated using google maps as grand truth. The error rate was calculated on average as 7.8% for the calculation of time and 17.2% for the calculation of number of transfers.

### 5.2 Statistics of Datasets

The quantitative and qualitative properties of the gathered data is summarized in tables 1 and 2.

**Table 1: Quantitative Properties of Studied Datasets**

| City | No. Stops | No. Routes | No. of Connected Stop Pairs |
|---|---|---|---|
| San Francisco | 1,926 | 54 | 3,481 |
| Boston | 4721 | 71 | 4,185 |
| Toronto | 5,806 | 144 | 9,594 |
| Los Angeles | 8431 | 108 | 12,240 |

**Table 2: Qualitative Properties of Studied Datasets**

| City | Stops | Routes | Connected Stop Pairs |
|---|---|---|---|
| Area | No. | Name/Number | Shared Routes |
| Population | Address | From Stop No. | Straight Distance |
|  | Latitude | To Stop No. | Road Distance |
|  | Longitude |  |  |

### 5.3 Visualizations and Results

Figure 1 depicts how a modeled transit network is overlaid on the satellite view of a city from Google Maps. For scaling, the locations of a few random stops are matched and the rest of the map is scaled accordingly.

In order to illustrate how the concepts in graph theory are meaningful in the transit network systems, consider the bridges, a.k.a. the routes that removing them would divide the model into two separate components (Figure 3).

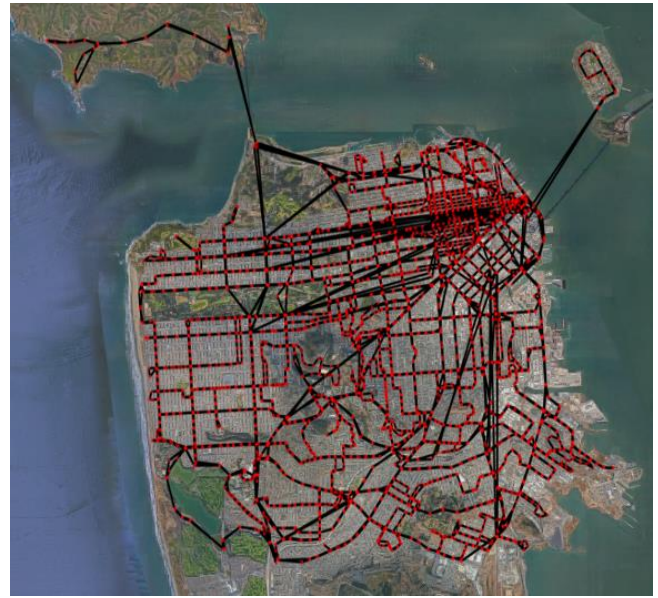

**Figure 1: Created model for the city of San Francisco.**

**Table 3: Total City Metric Results**

| Metric | San Francisco | Boston | Toronto | Los Angeles |
|---|---|---|---|---|
| Total Length (km) | 853 | 1064 | 2844 | 4838 |
| Total Travel Time (hour) | 3235 | 2300 | 7129 | 13285 |
| Speed (Km/h) | 15.8 | 27.6 | 23.9 | 21.9 |

**Table 4: Average Coverage Metric Results (Based on Area)**

| Metric | San Francisco | Boston | Toronto | Los Angeles |
|---|---|---|---|---|
| Trip Time (min) | 64 | 134 | 88 | 184 |
| Trip Time (per Km) (min) | 9.1 | 8.4 | 5.5 | 6.7 |
| Trip Length (Km) | 17 | 35 | 38 | 65 |
| Num. of Transfers/Trip | 1.5 | 2.1 | 2.6 | 3.2 |
| Num. of Transfers/Trip (per Km) | 0.22 | 0.14 | 0.16 | 0.12 |
| Straight Distance (Km) | 7 | 15.9 | 16 | 27.2 |
| Trip Length Ratio | 2.48 | 2.18 | 2.39 | 2.4 |
| Num. of stops within 400m | 5.6 | 2.4 | 3.1 | 2 |
| Distance to closest stop (m) | 245 | 334 | 260 | 331 |

**Table 5: Average Coverage Metric Results (Based on Population)**

| Metric | San Francisco | Boston | Toronto | Los Angeles |
|---|---|---|---|---|
| Trip Time (min) | 55.3 | 95.7 | 83.4 | 111.7 |
| Trip Time (per Km) (min) | 11 | 13.6 | 6.4 | 8.7 |
| Trip Length (Km) | 13 | 21.1 | 32 | 39.7 |
| Num. of Transfers/Trip | 0.3 | 0.24 | 0.2 | 0.12 |
| Num. of Transfers/Trip (per Km) | 5 | 7 | 13 | 12.8 |
| Straight Distance (Km) | 2.57 | 3 | 2.46 | 3.1 |
| Trip Length Ratio | 9 | 4.6 | 3.9 | 2.2 |
| Num. of stops within 400m | 133 | 220 | 207 | 300 |
| Distance to closest stop (m) | 0.3 | 0.24 | 0.2 | 0.12 |

**Table 6: Average Per Connection Metric Results**

| Metric | San Francisco | Boston | Toronto | Los Angeles |
|---|---|---|---|---|
| Trip Time (sec) | 56 | 33 | 45 | 65 |
| Trip Length (Km) | 245 | 254 | 296 | 395 |
| Wait Time (min) | 8.8 | 21.5 | 8.8 | 19.2 |
| Standard Deviation (min) | 4.6 | 9 | 4.8 | 7.3 |

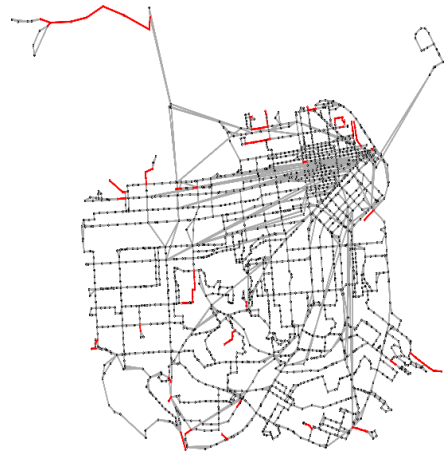

**Figure 2: Bridge connections for the San Francisco model.**

## 6 CONCLUSIONS

### 6.1 Highlights and Discussion

Many features can be extracted after the public transport system data has been represented as a graph of stops and connections.

There are evident features, such as network bridges: edges in the graph where, if they were to be removed, the graph would be disconnected, split into two or more subgraphs (Figure 2). In most cities, small roads leading to separated communities tend to be bridge edges. The actual bridges in San Francisco, however, are composed of two lanes-connections and therefore are not bridges.

Other features however, are not as apparent. The total length corresponds to the total length of metro/streetcar tracks and bus routes in the city, and is more for larger cities, as expected. However, the total travel time is not increased accordingly. As seen by the result of average speed, relatively San Francisco has higher speed and lower total time, compared to a smaller city like San Francisco. (Table 3). Average per connection, trip time and trip length, follow the same trend of the total system. On the other hand, it is well seen from (Table 4) that Boston and Los Angeles have much higher waiting times and also the standard deviation of them are almost twice Toronto and San Francisco.

According to tables 5 and 6, trip times in Boston and Los Angeles drop faster when calculated based on population density maps compared to when calculated based on a uniform area.

When normalized per kilometer, the trip time for Boston increases too fast. Its reverse happens with the metric on number of transfers from one point in the map to another for Los Angeles. This should mean that the structure of the public transit system in Boston and Los Angeles align well to the roads in the city and fewer public transit routes cover longer distances. However, when considering the population dense locations, Boston is not connected well. Also, Los Angeles has routes planned to connect such locations with less transfers compared to the average of the city.

The increase in number of stops within 400 meters walking distance and the decrease in distance to closest stop are visible again when the values are calculated based on population density maps. This is expected and means that public transit system is providing a better service in population dense areas compared to the average of the city. This is more accentuated in San Francisco and Boston. Which from another point of view could also mean that not all the city is well serviced by the public transit system.

## 6.2 Future Works

*6.2.1.* Population density is currently based on residential population density data of cities. However, to be more accurate, it is appealing to account for population density of workers and students during working hours when considering the efficiency of the transit system's access metrics.

*6.2.2. Graph theory concepts.* During this study, many concepts and properties in graph theory were mapped to the domain of public transit systems. As the study became more in depth, most algorithmic components were developed with domain specific intentions. The authors are interested to compare these algorithmic components which resulted in computation of the metrics in the domain with their general underlying concepts in graph theory.